\documentclass[a4paper,11pt]{article}
\usepackage{pos}
\usepackage{subcaption}
\usepackage{float}

\title{Energy-dependent polarization of Gamma-Ray Bursts' prompt emission with the POLAR and POLAR-2 instruments}
\ShortTitle{GRB energy-dependent polarization analysis with POLAR and POLAR-2}

\ConferenceLogo{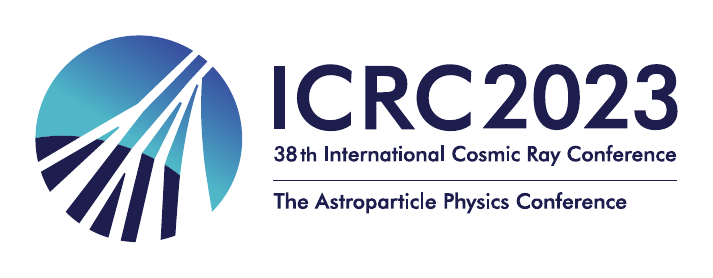}

\graphicspath{{Pictures}}

\author*[a]{Nicolas De Angelis}

\onbehalf{for the POLAR and POLAR-2 collaborations} 

\affiliation[a]{DPNC, University of Geneva,\\
  24 quai Ernest Ansermet, Geneva, Switzerland}

\emailAdd{nicolas.deangelis@unige.ch}
\emailAdd{nicolas.de.angelis@cern.ch}

\abstract{Gamma-Ray Bursts are among the most powerful events in the Universe. Despite half a century of observations of these transient sources, many open questions remain about their nature. Polarization measurements of the GRB prompt emission have long been theorized to be able to answer most of these questions.

With the aim of characterizing the polarization of these prompt emissions, a compact Compton polarimeter, called POLAR, has been launched to space in September 2016. Time integrated polarization analysis of the POLAR GRB catalog have shown that the prompt emission is lowly polarized or fully unpolarized. However, time resolved analysis depicted strong hints of an evolving polarization angle within single pulses, washing out the polarization degree in time integrated analyses. 

Here we will for the first time present energy resolved polarization measurements with the POLAR data. The novel analysis, performed on several GRBs, will provide new insights and alter our understanding of GRB polarization. The analysis was performed using the 3ML framework to fit polarization parameters versus energy in parallel to the spectral parameters. Although limited by statistics, the results could provide a very relevant input to disentangle between existing theoretical models.

In order to gather more statistics per GRB and perform joint time and energy resolved analysis, a successor instrument, called POLAR-2, is under development with a launch window early 2025 to the CSS. After presenting the first energy resolved polarization results of the POLAR mission, we will present the prospects for such measurements with the upcoming POLAR-2 mission.}

\FullConference{%
38th International Cosmic Ray Conference (ICRC2023)\\
  26 July - 3 August, 2023\\
  Nagoya, Japan}

\begin{document}
\maketitle

\section{The POLAR instrument's past results and the need for energy resolved polarization measurement of GRBs' prompt emission}

After several decade of spectral and temporal measurements, the emission mechanisms at play in Gamma-Ray Bursts (GRBs) as well as the magnetic field and jet structure still remain poorly understood. Polarization measurements have been theorized to be able to bring crucial information to disentangle between the existing models. As the first GRB dedicated $\gamma$-ray polarimeter, POLAR brought a new insight into the field of GRB polarization by measuring the polarization degree and angle for 14 sources. The instrument, divided into 25 modules each made of 64 channels \cite{Produit+18}, made use of Compton scattering to measure the polarization of the prompt emission of GRBs. Using the fact that a photon would preferentially scatter 90$^\circ$ with respect to its polarization vector, this array of 40$\times$40 plastic scintillator bars read out by multi-anode photomultiplier tubes is able to collect the scattering direction of the incoming $\gamma$-rays. By analyzing the scattering angle distribution the polarization degree (PD) and angle (PA) can be retrieved, respectively as the relative amplitude and the phase of the obtained sinusoidal modulation following the Klein-Nishina cross section.\\

Launched in low Earth orbit in September 2016 as a payload on the Tiangong-2 Chinese space laboratory, POLAR detected 55 GRBs as well as several solar flares all along its 6 months operation time. Polarization being a statistically demanding measurement, only 14 of these bursts were bright enough to deliver enough photons for a polarization analysis. The energy integrated analysis of these 14 GRBs have shown a tendency for low polarization fractions, even compatible with no polarization at all \cite{Kole+20}. By analyzing the time dependence of polarization for the brightest GRBs detected\footnote{Since performing a time resolved analysis requires to split the prompt emission into several time bins, this analysis was only possible on the few brightest detected sources.}, a quickly evolving polarization angle with a higher polarization degree has been hinted \cite{Burgess+19}. This result is compatible with the previously obtained low polarization degree on the integrated analysis, since the quick evolution of the PA would wash out the PD.\\

In addition to time resolved analysis of the polarization, another interesting measurement in order to gather more information on GRBs is the energy dependence of the polarization degree and angle \cite{Lundman+18}. We therefore present here an analysis performed on the POLAR GRB catalog \cite{Kole+20} to study the energy dependence of the polarization degree using a Heaviside function. Several other energy dependent functions are currently being investigated to fit the PD and PA as a function of energy, but this will be discussed in a later work \cite{NDA+23}. One of the main outcome from the POLAR experiment except for its science results is the need for a next generation of more sensitive $\gamma$-ray polarimeters in order to precisely measure time and energy resolved polarization. A bigger and more sensitive version of POLAR, called POLAR-2 \cite{Kole+ICRC21, NDA+ICRC21, Produit+ICRC23}, is therefore under development. A quick overview of this instrument will be given, while its sensitivity to energy dependence of polarization will be studied in a future paper \cite{NDA+23}.

\newpage
\section{Energy resolved polarization analysis with the POLAR data}

\begin{figure}[H]
\centering
     \hspace*{-3cm}\begin{subfigure}[b]{0.38\textwidth}
         \centering
         \includegraphics[height=\textwidth]{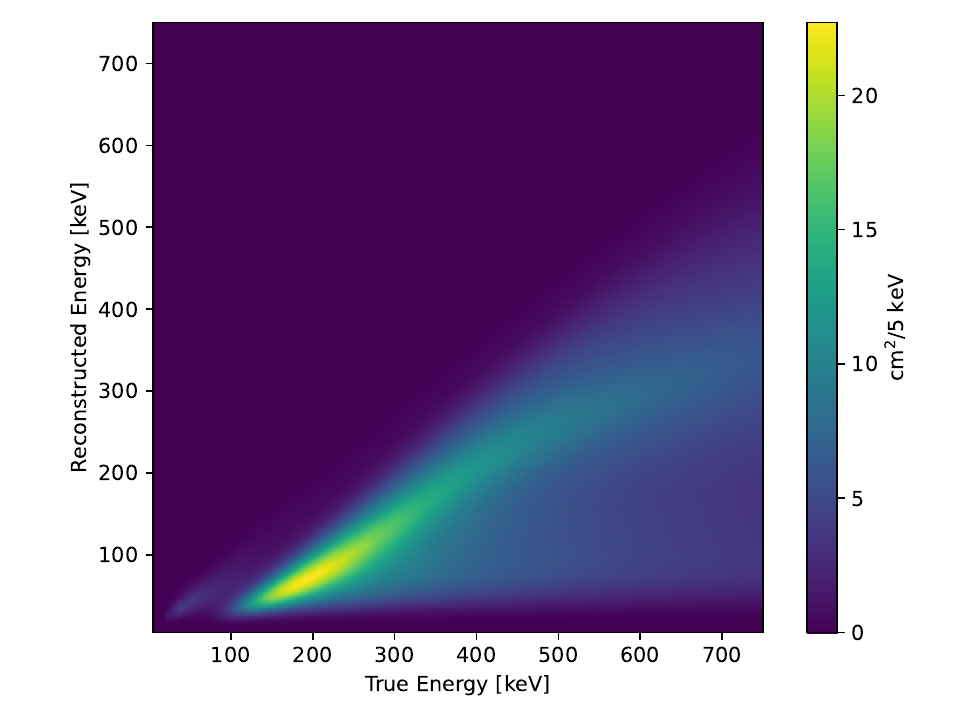}
         \caption{}
         \label{fig:energy_rsp_matrix}
     \end{subfigure}\hspace*{1.5cm}
    \begin{subfigure}[b]{0.38\textwidth}
         \centering
         \includegraphics[height=\textwidth]{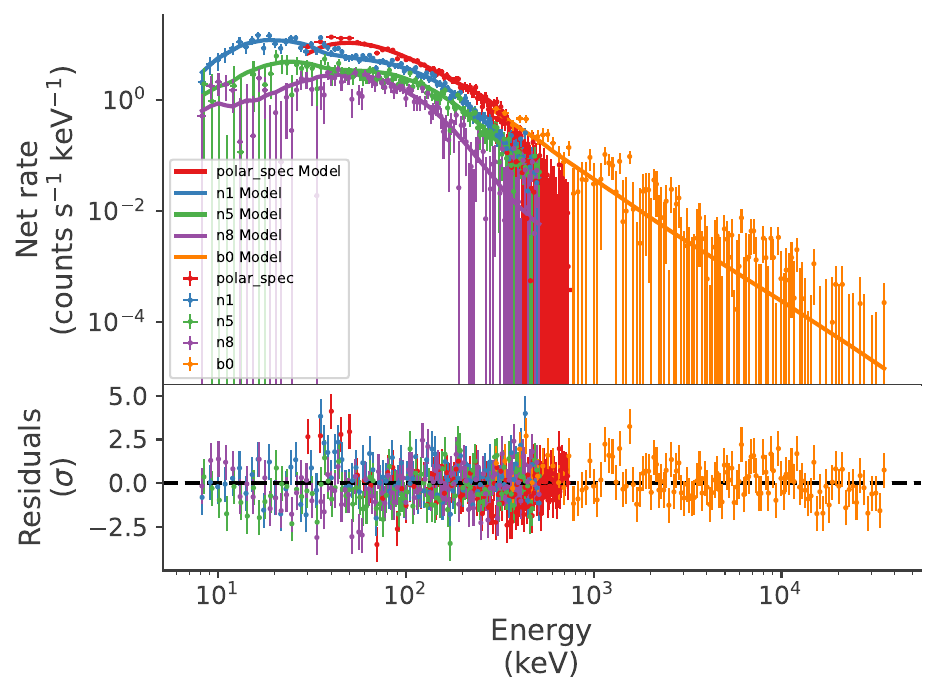}
         \caption{}
         \label{fig:170114A_spec}
     \end{subfigure}
 \caption{\textbf{a)} POLAR energy response matrix for GRB170114A. \textbf{b)} Spectral fit for POLAR and Fermi-GBM detectors based on the Band function.}
\end{figure}

POLAR is based on plastic scintillators in order to maximise the probability for Compton scattering, especially on the low energy side (few tens of keV). A drawback of using plastic based scintillating materials is the poor energy resolution, as shown in Figure \ref{fig:energy_rsp_matrix} where one can see a quite dramatic energy dispersion in the spectral response of the instrument. This will lead into bigger systematic errors in the resulting posterior distributions, but does not prevent us in any manner to perform energy dependent analysis since this energy spread is accounted for in the analysis using forward folding. The polarization and spectral components are fitted in parallel using the 3ML analysis framework \cite{3ML+ICRC15}, which also allow to perform joint fitting with other instruments' data. The spectral part is therefore fitted jointly with Fermi-GBM or Swift-BAT data when available. More details on the analysis can be found in \cite{Kole+20}.

The brightest GRB observed by POLAR is GRB170114A, for which we present here the energy integrated result already published in \cite{Kole+20}. The spectral component is fitted using the following Band function \cite{Band+93}:\\

$N_E(E) = K \begin{cases} \left(\frac{E}{100~keV}\right)^{\alpha} \exp \left(-\frac{(2+\alpha) E}{x_{p}}\right) & E \leq (\alpha-\beta) \frac{x_{p}}{(\alpha+2)} \\ \left(\frac{E}{100~keV}\right)^{\beta} \exp (\beta-\alpha)\left[\frac{(\alpha-\beta) x_{p}}{100~keV(2+\alpha)}\right]^{\alpha-\beta} &E>(\alpha-\beta) \frac{x_{p}}{(\alpha+2)} \end{cases} $\\

\begin{figure}[H]
\centering
\hspace*{-1cm}\includegraphics[height=0.3\textwidth]{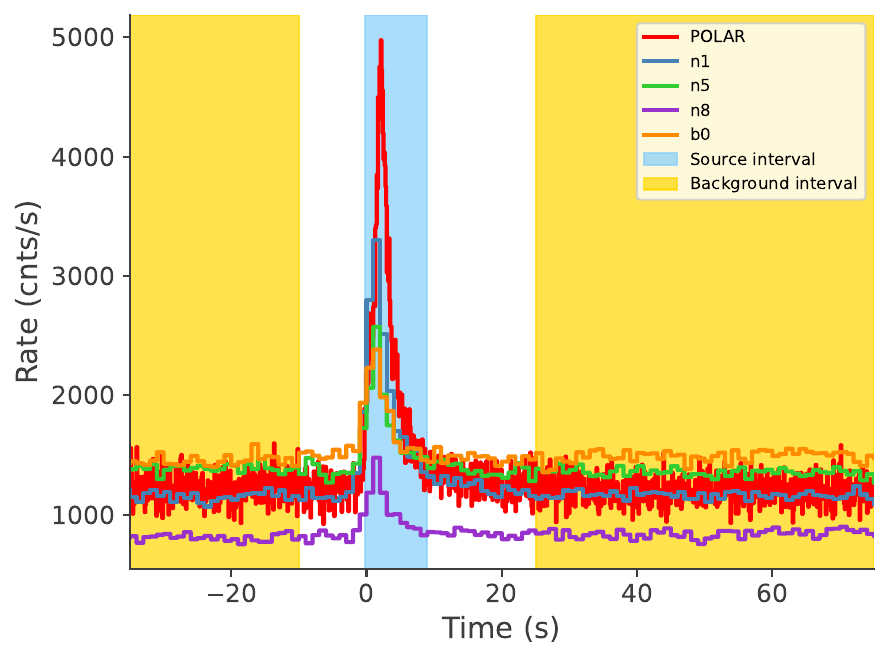}\includegraphics[height=0.3\textwidth]{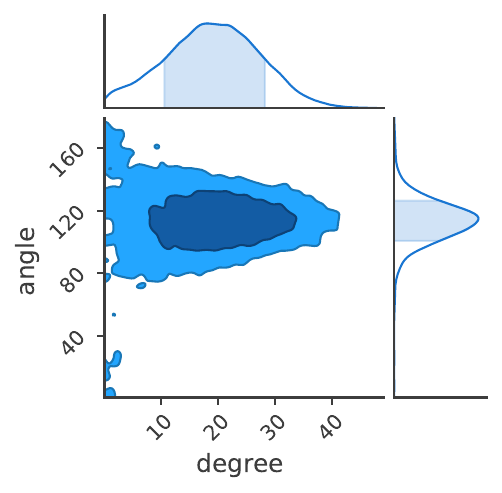}\includegraphics[height=0.3\textwidth]{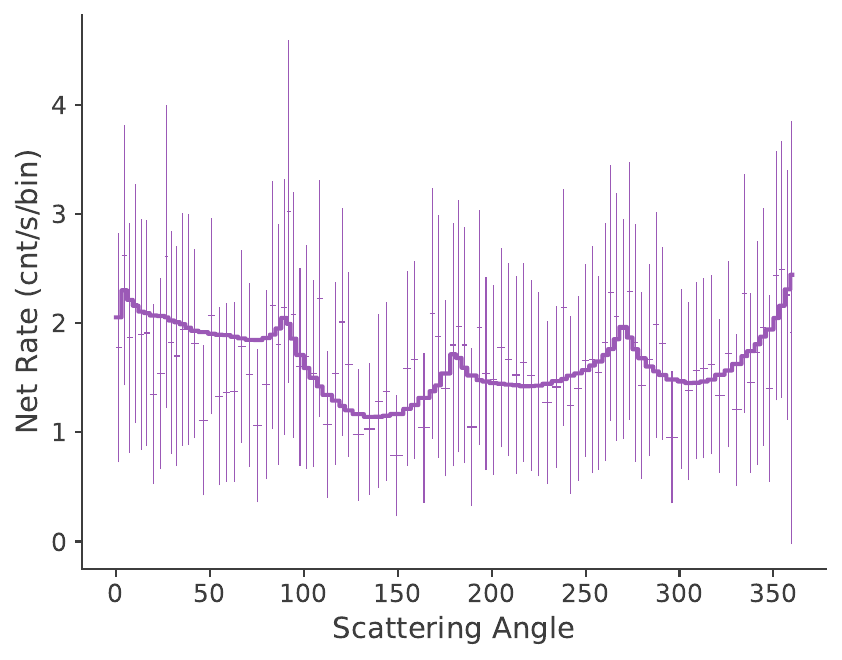}
 \caption{Light curves as seen by POLAR and the relevant GBM detectors for GRB170114A with background and source time intervals (left), resulting PD and PA values from the energy integrated analysis (middle), and corresponding fitted scattering angle distribution (right)}
 \label{fig:energy_integrated_170114A}
\end{figure}

The resulting parameters for the Band fit for the POLAR data of GRB170114A shown in Figure \ref{fig:170114A_spec} are: $K=4.56_{-0.25}^{+0.29}\cdot 10^{-2}$ keV$^{-1}$s$^{-1}$cm$^{-2}$, $\alpha=(-7.9\pm0.5)\cdot 10^{-1}$, $\beta=-2.31_{-0.11}^{+0.13}$, and $x_p=2.63_{-0.16}^{+0.18}\cdot 10^2$ keV. Figure \ref{fig:170114A_spec} also shows the spectral fit for the joint NaI and BGO detectors from Fermi-GBM. Note that although most of the spectral fits for the GRBs in the POLAR catalog are performed using the Band function, a few of them are fitted with a cutoff powerlaw (CPL), as described in \cite{Kole+20}. The fitted scattering angle distribution as well as the resulting PD and PA are shown in figure \ref{fig:energy_integrated_170114A}, while the obtained values for the polarization parameters are shown in the first column of Table \ref{table:pol_res}. Figure \ref{fig:energy_integrated_170114A} also shows the light curves of all the detectors used for the analysis of GRB170114A as well as the time selections for the source and background.

\begin{figure}[H]
     \centering
     \begin{subfigure}[b]{0.49\textwidth}
         \centering
         \includegraphics[height=\textwidth]{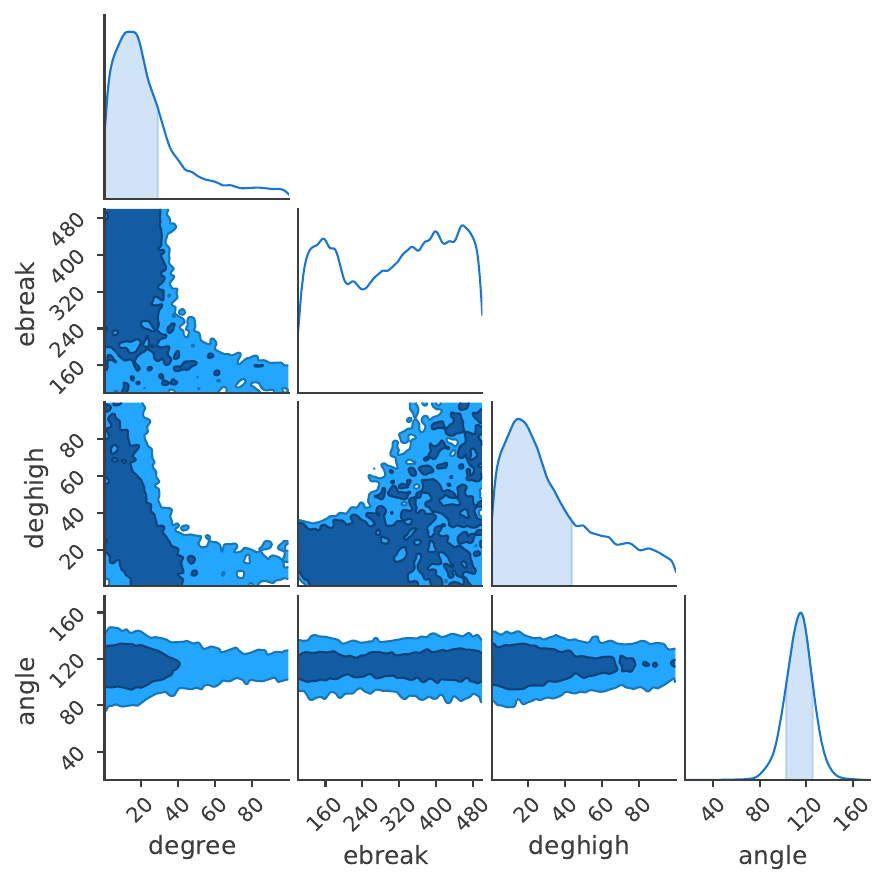}
         \caption{}
         \label{fig:pdheayside_corner}
     \end{subfigure}
    \begin{subfigure}[b]{0.49\textwidth}
         \centering
         \includegraphics[height=\textwidth]{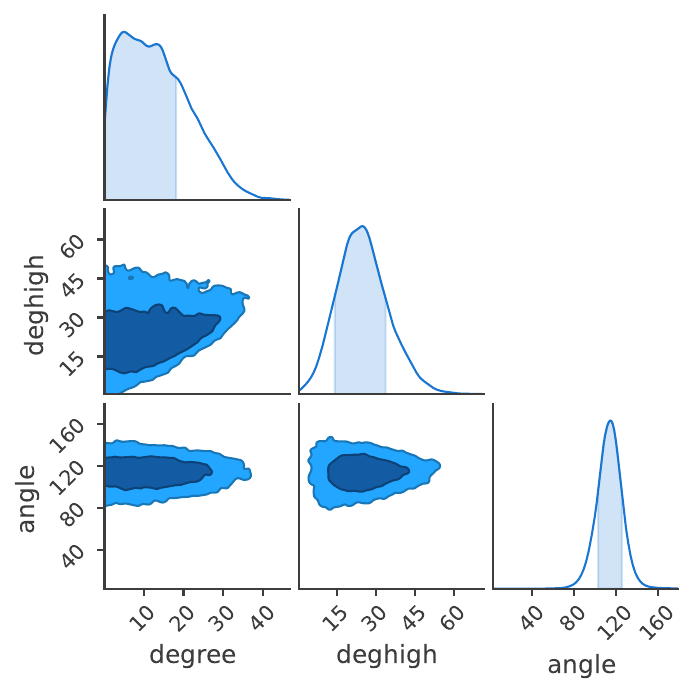}
         \caption{}
         \label{fig:paheavyside_corner}
     \end{subfigure}
        \caption{Posterior distributions of polarization parameters for different energy resolved models: \textbf{a)} Heaviside fit on the polarization degree. The 3 polarization parameters are the PD at low and high energies and the energy break separating the two regions. This latter parameter is unconstrained while the polarization degrees at low and high energy are compatible to each other. \textbf{b)} Heaviside fit on the polarization degree with a fixed energy break at 150~keV. The 2 polarization parameters are the PD at low and high energies. No significant variation of the PD with energy is observed.}
        \label{fig:corner_plots}
\end{figure}

The polarization degree has been fitted for GRB170114A using a Heaviside function, while keeping the PA constant with energy. Two trials have been made, one with a free energy break (position of the "step") and one by fixing the energy break to 150~keV. In the former case, the 4 polarization parameters are the PD at low and high energy, the PA, and the energy break position, while in the latter case only 3 polarization parameters are used since the energy break is fixed. The resulting "corner plots" for both cases, showing the posterior distributions of each parameter against each other, are shown in Figure \ref{fig:corner_plots}.


\begin{table}[H]
\centering
\begin{tabular}{lccc} 
 \hline
  & Energy integrated & PD\textsubscript{Heaviside} with a free break & PD\textsubscript{Heaviside} for E\textsubscript{break}=150keV \\ [0.5em] 
 \hline\hline
 PD [\%] & $22.8_{-12.3}^{+45.5}$ & - & - \\ [0.2em]
 PD\textsubscript{low} [\%] & - & $5.71_{-1.20}^{+35.69}$ & $19.6_{-15.5}^{+2.7}$ \\ [0.2em]
 PD\textsubscript{high} [\%] & - & $79.6_{-70.0}^{+14.4}$  & $19.5_{-4.0}^{+15.3}$ \\ [0.2em]
 PA [$^\circ$] & $117_{-16}^{+9}$ & $115_{-12}^{+10}$ & $109_{-7}^{+15}$ \\ [0.2em]
 E\textsubscript{break} [keV] & - & $334_{-172}^{+111}$ & 150 \\ [0.5em] 
 \hline
\end{tabular}
\caption{Mean and standard deviation of the polarization parameters for different fits (energy integrated, Heaviside polarization degree with a free or fixed energy break) extracted from the posterior distributions shown in Figures \ref{fig:energy_integrated_170114A} and \ref{fig:corner_plots}.}
\label{table:pol_res}
\end{table}

\begin{figure}[H]
\centering
     \hspace*{-2cm}\begin{subfigure}[b]{0.55\textwidth}
         \centering
         \includegraphics[height=\textwidth]{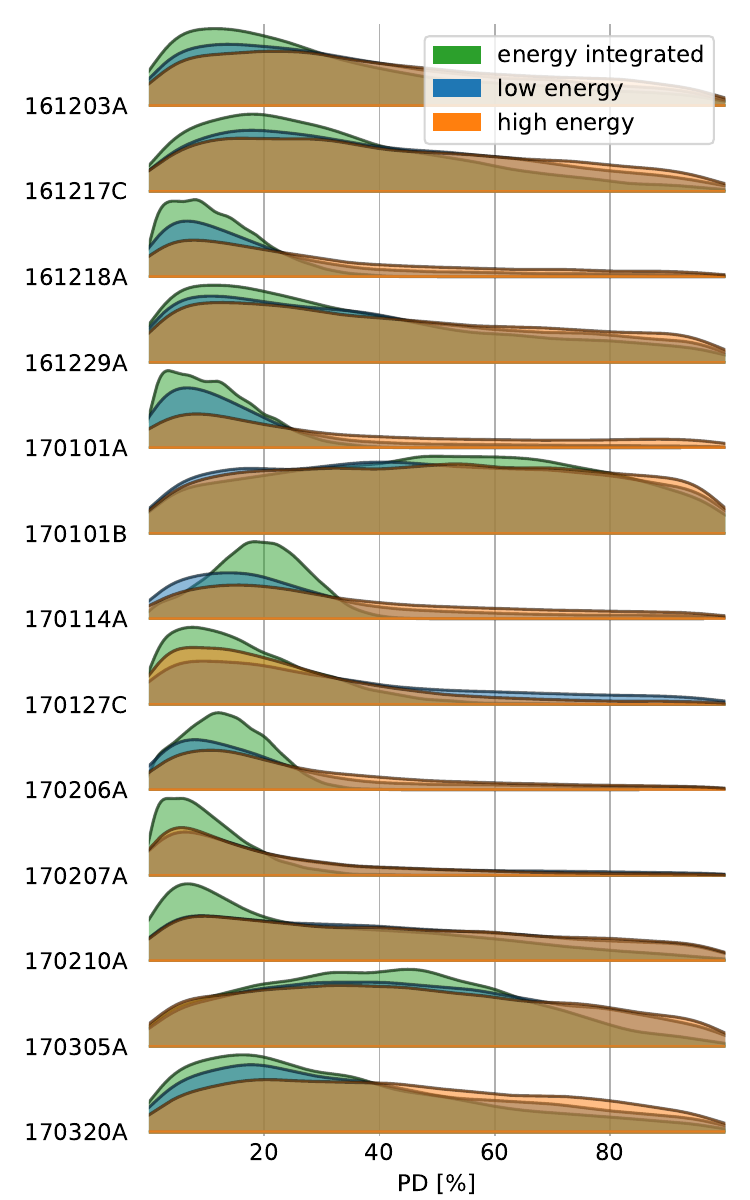}
         \caption{}
         \label{fig:pdheayside_PD_catalog}
     \end{subfigure}\hspace*{-3cm}
    \begin{subfigure}[b]{0.55\textwidth}
         \centering
         \includegraphics[height=\textwidth]{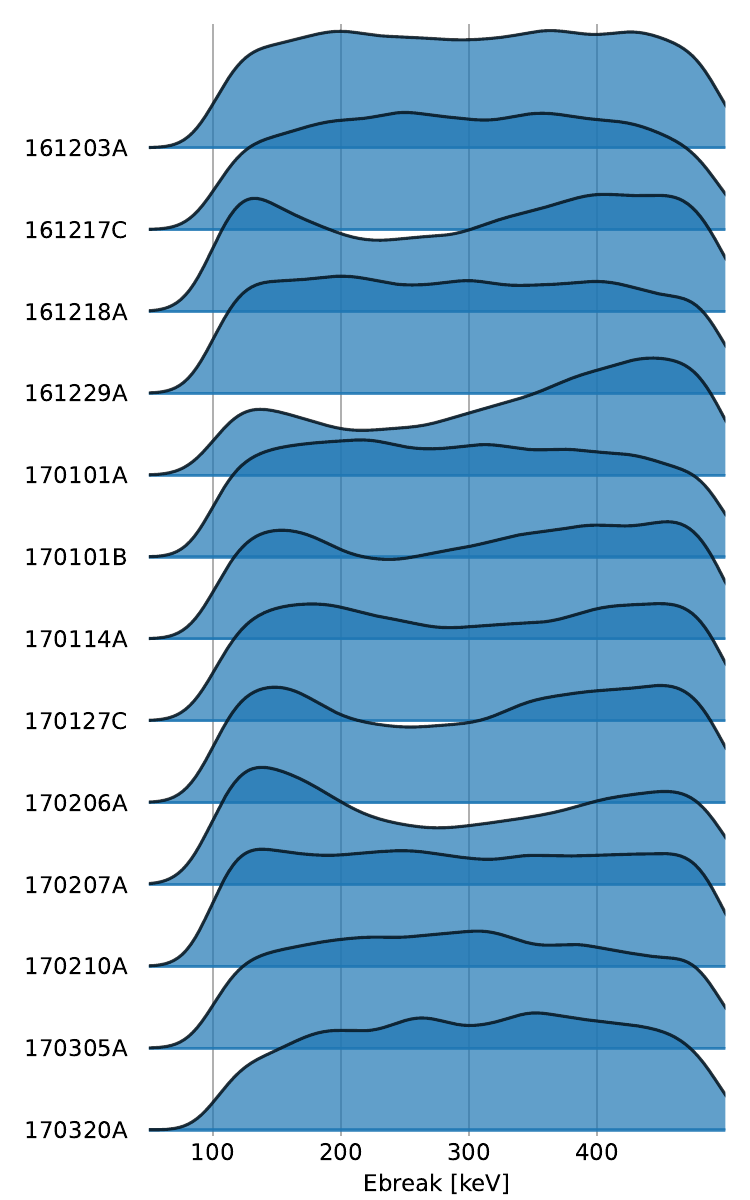}
         \caption{}
         \label{fig:pdheavyside_Ebreak_catalog}
     \end{subfigure}\hspace*{-3cm}
    \begin{subfigure}[b]{0.55\textwidth}
         \centering
         \includegraphics[height=\textwidth]{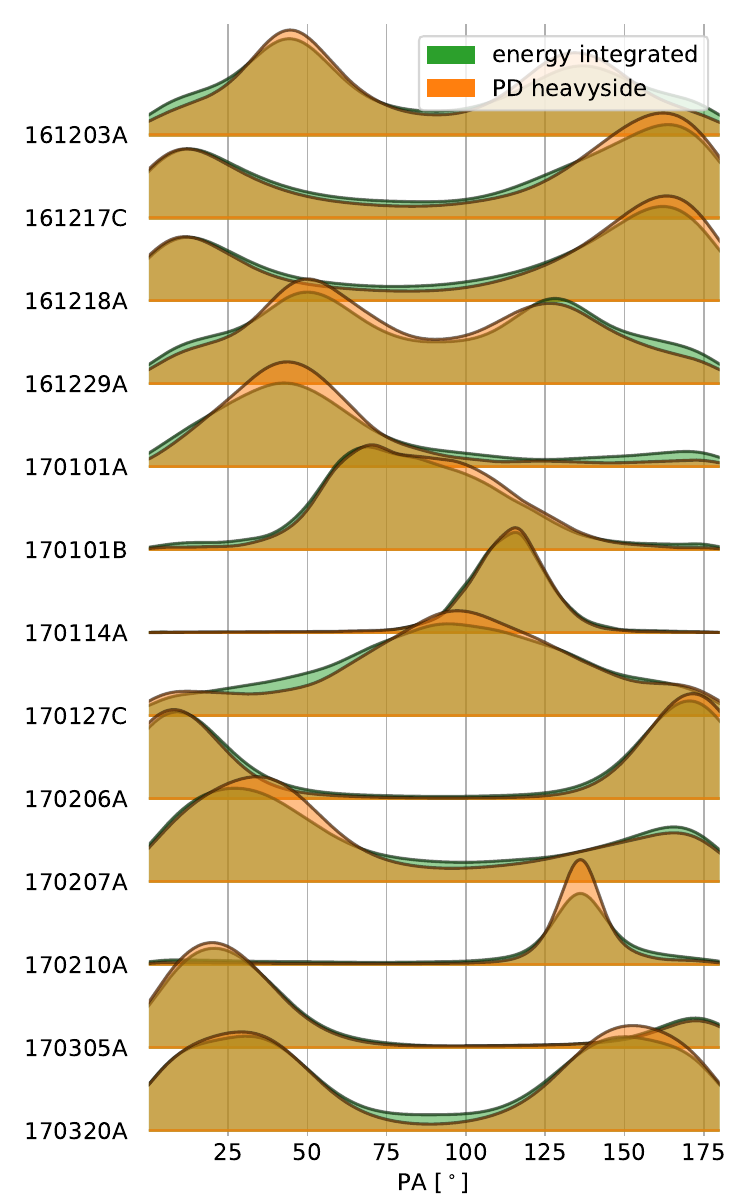}
         \caption{}
         \label{fig:pdheavyside_PA_catalog}
     \end{subfigure}
 \caption{Resulting polarization parameters' posterior distributions for the PD Heaviside fit on the POLAR GRB catalog compared to what was obtained in the energy integrated case \cite{Kole+20}. \textbf{a)} Polarization degree for the integrated case and at low and high energy for the Heaviside fit. The three PD values are compatible with each other for all the GRBs, the PD posterior distribution for the energy integrated case being slightly more constrained due to higher statistics, since the events in the case of the PD Heaviside fit are split into 2 energy bins. \textbf{b)} Energy break for the Heaviside fit on the PD. For most of the GRBs it is poorly constrained except at very low energies below 100~keV \textbf{c)} Polarization angle in the case of an Heaviside PD fit and in the energy integrated case. The PA for both cases are compatible for all GRBs of the catalog.}
 \label{pdheavyside_catalog}
\end{figure}

\newpage
The obtained values for the polarization parameters for both energy dependent fits are compared to those obtained in the energy integrated fit in Table \ref{table:pol_res}. The polarization angle is very similar in the three cases. In the case where the energy break is free, we can see that the position of the break is poorly constrained, so as the PD at high energy, while the PD at low energy is compatible with a low polarization level. When the break is fixed to 150~keV, both PDs at low and high energy are compatible with each other and with the energy integrated result. By looking at the posterior distributions of both PDs, the PD at low energy seems a bit more constrained to low values while the PD at high energy is spanning higher values.

These two fitting functions have been used to analyse the entire POLAR catalog. Figure \ref{fig:pdheayside_PD_catalog} shows the posterior distributions for the PD in the energy integrated compared to the PDs at low and high energy in the case where the energy break is free. For all the GRBs the three PD values are compatible with each other, while PD\textsubscript{low} and PD\textsubscript{high} are a bit less constrained due to lower statistics, since fitting the data with a Heaviside function is equivalent two dividing the events into 2 energy bins. The energy break, shown in Figure \ref{fig:pdheavyside_Ebreak_catalog}, is only constrained at low energy as we can see that values below 100~keV are highly suppressed. The PA for the energy integrated and resolved cases, as for GRB170114A, are compatible, as depicted in Figure \ref{fig:pdheavyside_PA_catalog}.

\begin{figure}[H]
\centering
\includegraphics[height=0.6\textwidth]{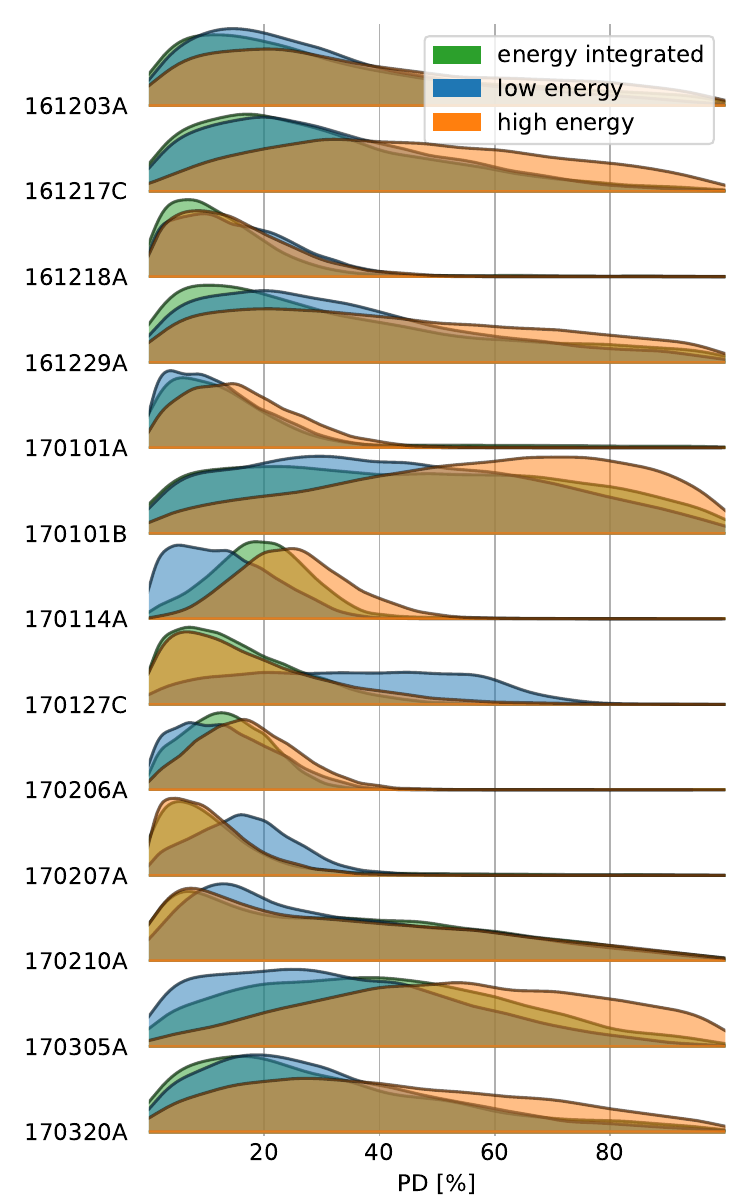}\hspace*{0.5cm}\includegraphics[height=0.6\textwidth]{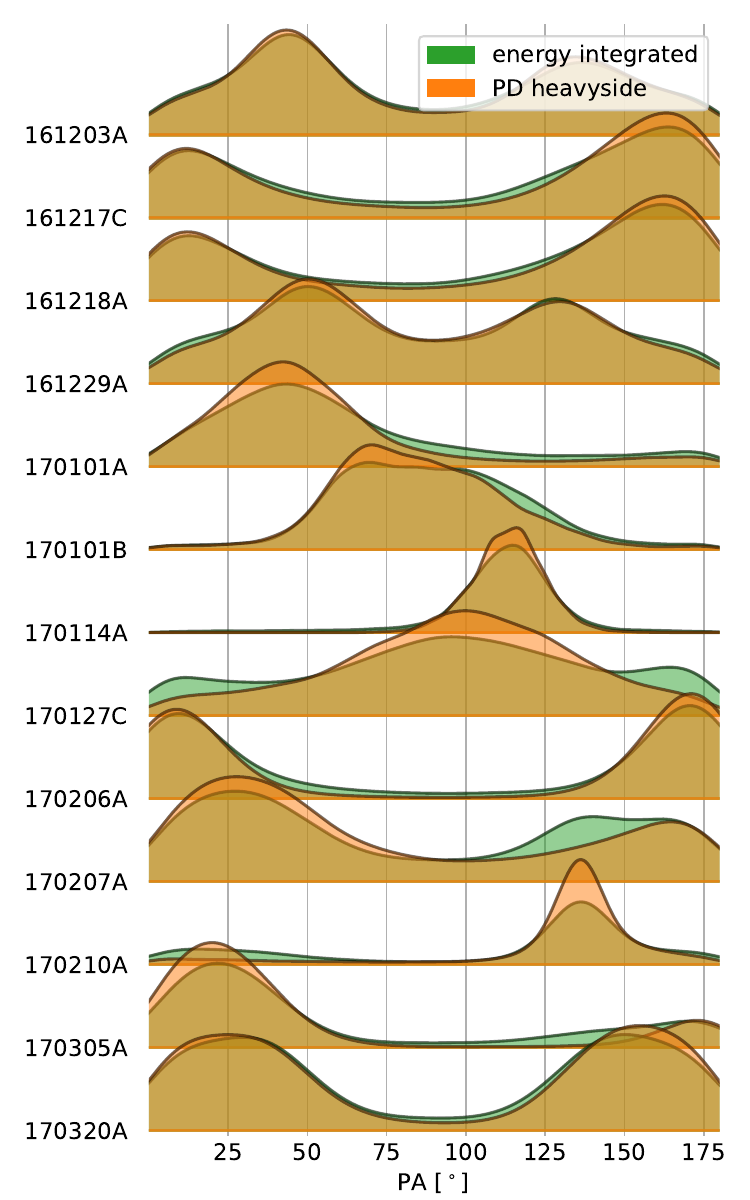}
 \caption{Result of the PD Heaviside fit for a fixed 150~keV energy break on the POLAR GRB catalog compared to the energy integrated results \cite{Kole+20}. \textbf{Left:} Polarization degree for the integrated case and at low and high energy for the Heaviside fit. The three PD values are compatible with each other for all the GRBs, although there seems to be a difference between the PD at low and high energy in the posterior distributions. The limited statistics is preventing a more precise resolution between these two PD values. \textbf{Right:} Comparison of the obtained PA for the energy integrated and resolved cases. The PA for both cases are compatible for all GRBs of the catalog.}
 \label{fig:pdheavyside_fixedbreak150_catalog}
\end{figure}

\newpage
By fixing the energy break, we still obtain a very similar results than in the energy integrated case for the PA. By checking the values of the PD, no significant energy dependence is observed. By looking at the posterior distributions of the PD (see Figure \ref{fig:pdheavyside_fixedbreak150_catalog}), the PD at low energy seems again more constrained towards low values than the PD at high energy, which is spanning towards higher polarization fraction values.\\

By using a Heaviside fit, no significant energy dependence of the PD is observed on the POLAR catalog and the fitting parameters are not well constrained due to limited statistics and energy resolution. The next step to this work is to check for energy dependence of the polarization angle using a Heaviside function, and to try a linear fit of the PD and PA as a function of energy.

\section{Future prospects: the POLAR-2 instrument}

\begin{figure}[H]
\centering
     \hspace*{-2cm}\begin{subfigure}[b]{0.4\textwidth}
         \centering
         \includegraphics[height=\textwidth]{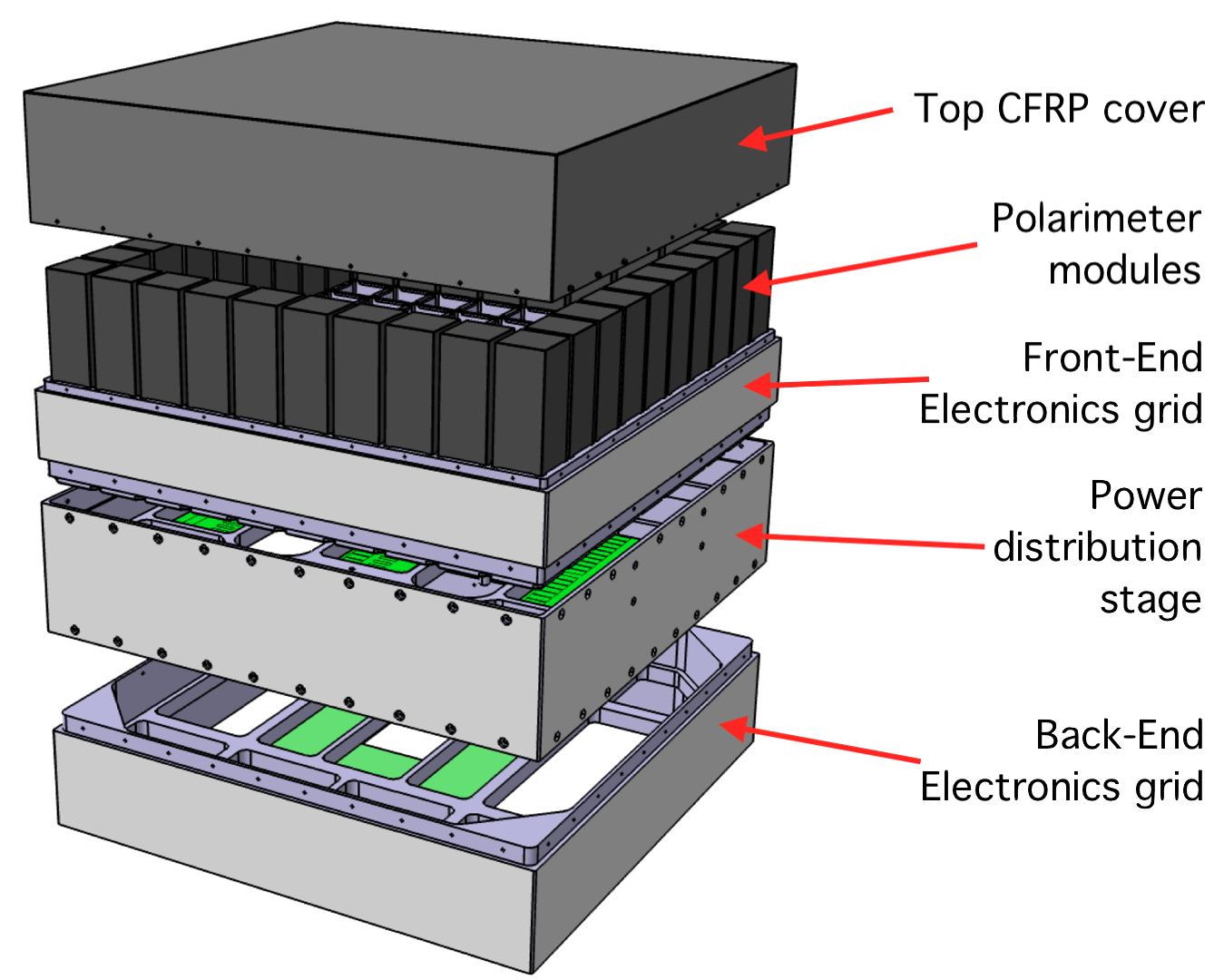}
         \caption{}
         \label{fig:POLAR-2_CAD}
     \end{subfigure}\hspace*{1.5cm}
    \begin{subfigure}[b]{0.4\textwidth}
         \centering
         \includegraphics[height=\textwidth]{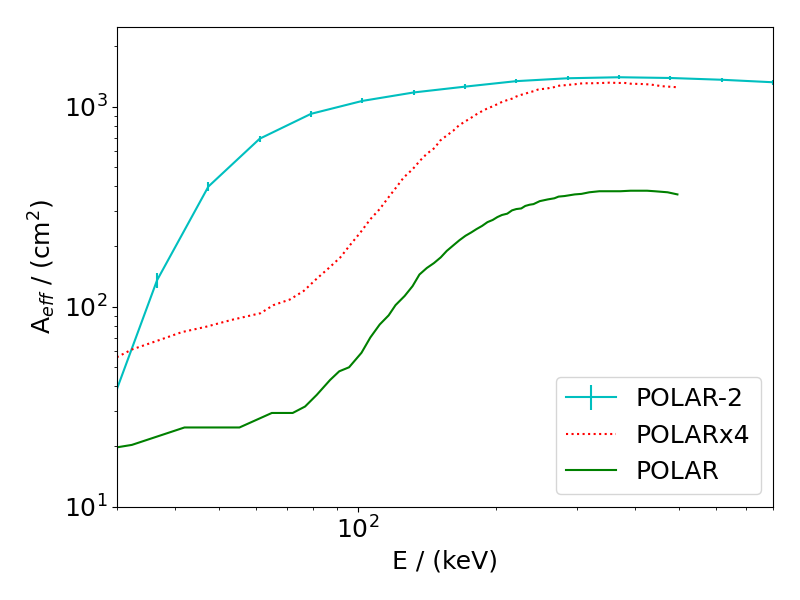}
         \caption{}
         \label{fig:POLAR-2_aeff}
     \end{subfigure}
     
 \caption{\textbf{a)} Exploded view of the POLAR-2 instrument design. The instrument is composed of three main mechanical grid: the one at the top, facing the deep space, contains the sensitive detector modules and their front-end electronics; the middle one is responsible for signal collection and power distribution to the 100 individual modules; and the bottom grid is responsible for power conversion from the space station to the polarimeter, communication with the space station, and is hosting the central computer of the payload. \textbf{b)} Effective area comparison between POLAR and POLAR-2. Thanks to technological upgrades improving the light yield of the polarimeter (amount of detected light per incoming energy), we see a considerable sensitivity improvement at low energies compared to the scaled up old POLAR design.}
 \label{POLAR-2}
\end{figure}

The statistics being one of the main limiting factor for precisely measuring the polarization of GRBs, a bigger Compton polarimeter based on the legacy of POLAR is under development to be launched to the China Space Station in 2025 \cite{Kole+ICRC21, NDA+ICRC21, Produit+ICRC23}. Called POLAR-2, this instrument will contain 4 times more channels, as depicted in the CAD model in Figure \ref{fig:POLAR-2_CAD}. Since the instrument is background dominated due to its wide field of view, needed to gather as many GRBs as possible, the scintillator bars have been shortened in order to optimize the signal-to-noise ratio without reducing too much the effective area. Silicon photomulipliers (SiPMs), much more sensitive than the MA-PMTs used in POLAR, are being used to read out the plastic scintillators, which allows a better sensitivity especially at low energy. This is shown in Figure \ref{fig:POLAR-2_aeff}, where the POLAR-2 effective area as a function of energy is compared to the effective area of 4 times POLAR, that is keeping the same design as POLAR without any technological improvements and just quadrupling the number of modules.\\

In addition to trying different energy dependent polarization functions for fitting the POLAR data as mentioned in the previous section, the next steps of this work is to compare the sensitivity to polarization's energy dependence between POLAR and POLAR-2 using both instruments' response. A higher sensitivity could allow to perform energy and time resolved analysis at the same time, or at least look for a correlation between temporal evolution of the spectrum and polarization with the energy dependence of polarization.

\newpage

\clearpage
\section*{Full Authors List: POLAR and POLAR-2 Collaborations}

%
\scriptsize
\noindent
Nicolas De Angelis$^1$, 
J. Michael Burgess$^2$, 
Franck Cadoux$^1$, 
Jochen Greiner$^2$, 
Merlin Kole$^1$, 
Hancheng Li$^3$, 
Slawomir Mianowski$^4$, 
Agnieszka Pollo$^4$, 
Nicolas Produit$^3$, 
Dominik Rybka$^4$, 
Jianchao Sun$^5$,
Xin Wu$^1$
and 
Shuang-Nan Zhang$^{5,6}$ \\

\noindent
$^1$University of Geneva, DPNC, 24 Quai Ernest-Ansermet, CH-1211 Geneva, Switzerland.\\
$^2$Max-Planck-Institut fur extraterrestrische Physik, Giessenbachstrasse 1, D-85748 Garching, Germany.\\
$^3$University of Geneva, Geneva Observatory, ISDC, 16, Chemin d’Ecogia, CH-1290 Versoix, Switzerland.\\
$^4$National Centre for Nuclear Research, ul. A. Soltana 7, 05-400 Otwock, Swierk, Poland.\\
$^5$Key Laboratory of Particle Astrophysics, Institute of High Energy Physics, Chinese Academy of Sciences, Beijing 100049, China.\\
$^6$University of Chinese Academy of Sciences, Beijing 100049, China.

\end{document}